\documentclass[times, 10pt,twocolumn]{article}
\usepackage{times}
\usepackage{amssymb,amsmath,latexsym}
\newtheorem{theorem}{Theorem}[section]
\newtheorem{lemma}[theorem]{Lemma}

\newtheorem{definition}[theorem]{Definition}
\newtheorem{example}{Example}

\newenvironment{proof}{\noindent
  \textbf{Proof.}}{\hfill$\Box$\\}

\newenvironment{proofsketch}{\noindent
  \textbf{Proof sketch.}}{\hfill$\Box$\\}

\newcommand{\instr}[5]{\ensuremath{\hbox to 60 pt
    {${#1}$\hfil${#2}$\hfil$ \rightarrow
      $\hfil${#3}$\hfil${#4}$\hfil${#5}$}}}

\newcommand{\cclass}[1]{\ensuremath{\mathbf{#1}}}
\newcommand{\bigo}[1]{\ensuremath{\mathcal{O}(#1)}}

\newcommand{\ecl}[1]{\ensuremath{\mathsf{ecl}(#1)}}

\newcommand{\D}{\ensuremath{\Delta}}
\newcommand{\G}{\ensuremath{\Gamma}}

\newcommand{\vp}{\ensuremath{\varphi}}

\newcommand{\fm}[1]{\emph{#1}}
\newcommand{\de}[1]{\emph{#1}}

\newcommand{\rel}[1]{\ensuremath{\mathcal{#1}}}

\newcommand{\power}[1]{\ensuremath{\mathcal{P}(#1)}}

\newcommand{\union}{\, \cup \,}

\newcommand{\bigunion}{\bigcup \,}
\newcommand{\biginter}{\bigcap \,}

\newcommand{\inter}{\, \cap \,}

\newcommand{\crh}[2]{\ensuremath{\{\, #1 \mid \, #2\, \}}}

\newcommand{\set}[1]{\ensuremath{ \{ #1 \} }}

\newcommand{\lang}{\ensuremath{\mathcal{L}}}

\newcommand{\system}[1]{\ensuremath{\mathbf{#1}}}

\newcommand{\LTL}{\ensuremath{\textbf{LTL}}}

\newcommand{\con}{\wedge}

\newcommand{\equivalence}{\leftrightarrow}

\newcommand{\ap}{\textbf{\texttt{AP}}}

          \newcommand{\until}{\ensuremath{\hspace{2pt}\mathcal{U}}}

\newcommand{\kframe}[1]{\ensuremath{\mathfrak{#1}}}

\newcommand{\mmodel}[1]{\ensuremath{\mathcal{#1}}}
\newcommand{\hintikka}[1]{\ensuremath{\mathcal{#1}}}

\newcommand{\sat}[3]{\ensuremath{\mmodel{#1}, #2 \Vdash #3}}

\newcommand{\cf}[1]{\ensuremath{\textsf{#1}}}

\newcommand{\cl}[1]{\ensuremath{\mathsf{cl}(#1)}}

\newcommand{\Rule}[1]{\textbf{(#1)}}

\newcommand{\subf}[1]{\ensuremath{\mathsf{Sub}(#1)}}

\newcommand{\ATL}{\textbf{ATL}}

\newcommand{\agents}{\ensuremath{\Sigma}}

\newcommand{\st}[1]{\ensuremath{\mathbf{states}(#1)}}

\newcommand{\brancharrow}{\ensuremath{\Longrightarrow}}

\newcommand{\tableau}[1]{\ensuremath{\mathcal{#1}}}

\newcommand{\knows}[1]{\ensuremath{\mathbf{K}}_{#1}}

\newcommand{\common}[1]{\ensuremath{\mathbf{C}}_{#1}}
\newcommand{\distrib}[1]{\ensuremath{\mathbf{D}}_{#1}}
\newcommand{\maelcd}{\textbf{MAEL}(CD)}

\newcommand{\psmodel}[1]{\ensuremath{\mathcal{M}^*}}
\newcommand{\pseudomodel}[1]{\ensuremath{\mathcal{M}^{**}}}

\newcommand{\cut}[1]{}

\begin{document}

\title{Tableau-based decision procedure for the multi-agent epistemic
  logic with operators of common and distributed knowledge}

\author{Valentin Goranko \\
  University of the Witwatersrand \\ School of Mathematics \\
  WITS 2050, Johannesburg, South Africa\\ goranko@maths.wits.ac.za\\
\and
Dmitry Shkatov\\
University of the Witwatersrand \\
School of Computer Science \\ WITS 2050, Johannesburg, South Africa\\
dmitry@cs.wits.ac.za\\
}
\date{}
\maketitle

\thispagestyle{empty}

\begin{abstract}
  We develop an incremental-tableau-based decision procedure for the
  multi-agent epistemic logic \maelcd\ (aka $S5_n (CD)$), whose
  language contains operators of individual knowledge for a finite set
  $\agents$ of agents, as well as operators of distributed and common
  knowledge among all agents in $\agents$. Our tableau procedure works
  in (deterministic) exponential time, thus establishing an upper
  bound for \maelcd-satisfiability that matches the (implicit)
  lower-bound known from earlier results, which implies
  \cclass{ExpTime}-completeness of \maelcd-satisfiability.  Therefore,
  our procedure provides a complexity-optimal algorithm for checking
  \maelcd-satisfiability, which, however, in most cases is much more
  efficient. We prove soundness and completeness of the procedure, and
  illustrate it with an example.
\end{abstract}

\section{Introduction}

Over the last two decades, multi-agent epistemic logics
(\cite{Fagin95knowledge, vdHM95}) have played a significant role in
computer science and artificial intelligence.  The main application
seems to have been to design, specification, and verification of
distributed protocols (\cite{HM90}), but a plethora of other
applications are described in, among others,~\cite{FHV92},
\cite{Fagin95knowledge} and \cite{vdHM95}.

Languages of multi-agent epistemic logics considered in the literature
contain various repertoires of modal operators.  In the present paper,
we consider the ``full'' multi-agent epistemic logic, which we call
\maelcd, whose language contains operators of individual knowledge for
a non-empty, finite set $\agents$ of agents as well as operators of
common (C) and distributed (D) knowledge among all agents in
$\agents$.  (Since all modal operators of \maelcd\ are
\system{S5}-modalities, the logic is also referred to in the
literature as $S5_n (CD)$).  To be used for such tasks as designing
protocols conforming to a given specification, \maelcd, needs to be
equipped with an algorithm checking for \maelcd-satisfiability.  The
first step in that direction was taken in~\cite{vdHM97}, where the
decidability of \maelcd\ has been established by showing that it has a
finite model property.  This result was proved in~\cite{vdHM97} via
filtration; therefore, the decision procedure suggested by that
argument is based on an essentially  brute-force enumeration of all finite models for \maelcd, which suggest a satisfiability-checking algorithm that is
theoretically important, but of limited practical value. Our tableau
procedure has, in comparison, the following advantages:

\begin{enumerate}
\item It establishes a (deterministic) \cclass{ExpTime} upper-bound
  for \maelcd-satisfiability, which matches the lower-bound that
  follows from the results of~\cite{HM92}.

\item It provides an algorithm for checking \maelcd-satisfiability
  that is not only provably complexity-optimal, but which in the vast
  majority of cases requires much less resources than what is
  predicted by the worst-case upper bound. This is one of the
  hallmarks of incremental tableaux (\cite{Wolper85}) as opposed to
  the top-down tableaux in the style of~\cite{EmHal85}, which
  \emph{always} require the amount of resources predicted by the
  worst-case complexity estimate. Top-down tableaux for the fragment
  of \maelcd\ not containing the operator of distributed knowledge
  have been presented in~\cite{HM92}.
\end{enumerate}

The type of incremental tableau developed herein originates
in~\cite{Wolper85}; tableaux in a similar style were recently
developed for the multi-agent logic \ATL\ and some of its variations
in~\cite{GorSh08}.  Thus, the present paper continues the enterprize
of designing complexity-optimal decision procedures for logics used in
design, specification and verification of multi-agent systems
(\cite{Fagin95knowledge, Wooldridge02}).  The particular style of the
tableaux presented here is meant to be compatible with the tableaux
from~\cite{GorSh08}, so that we can in the future build tableaux for
more sophisticated logics for multi-agent systems.

The main reason for the restriction of the distributed and common
knowledge operators only to be (implicitly) parameterized by the whole
set of agents referred to in the language, adopted in this paper, is
to be able to present the main ideas and features of the tableaux in
sufficient detail, while avoiding some additional technical
complications arising in the case of several such operators, each one
associated with a non-empty subset of the set of all agents.  This,
more complicated, case will be treated in a follow-up paper.

\section{Syntax and semantics of \maelcd}
\subsection{Syntax}

The language $\lang$ of \maelcd\ contains a (possibly,
countably-infinite) set \ap\ of \de{atomic propositions}, typically
denoted by $p, q, r, \ldots$; a finite, non-empty set $\agents$ of
(names of) \de{agents}, typically denoted by $a, b \ldots$; a
sufficient repertoire of the Boolean connectives; and the modal
operators $\knows{a}$ (``the agent $a$ knows that \ldots''),
$\distrib{}$ (``it is distributed knowledge among $\agents$ that
\ldots'') and $\common{}$ (``it is common knowledge among $\agents$
that \ldots''). Thus, the formulae of $\lang$ are defined as follows:
\[\vp := p \mid \neg (\vp) \mid (\vp_1 \con \vp_2) \mid \knows{a} (\vp)
\mid \distrib{} (\vp) \mid \common{} (\vp),\] where $p$ ranges over
$\ap$ and $a$ ranges over $\agents$.  The other boolean connectives
can be defined in the usual way.  We omit parentheses in formulae
whenever it does not result in ambiguity.  We denote arbitrary
formulae of $\lang$ by $\vp, \psi, \chi, \ldots$ (possibly with
decorations).  We write $\vp \in \lang$ to mean that $\vp$ is a
formula of $\lang$.  Formulae of the form $\neg \common{} \vp$ are
called \de{eventualities}.

\subsection{Semantics}

Formulae of $\lang$ are interpreted over multi-agent epistemic models,
based on multi-agent epistemic frames.  We will also need a more
general notion of multi-agent epistemic structure.

\begin{definition}
  \label{def:maes}
  A \de{multi-agent epistemic structure} (MAES, for short) is a tuple
  $\kframe{S} = (\agents, S, \set{\rel{R}_a}_{a \in \agents},
  \rel{R}_D, \rel{R}_C)$, where
  \begin{enumerate}
    \itemsep-4pt
  \item $\agents$ is a finite, non-empty set of agents;
  \item $S \ne \emptyset$ is a set of \de{states};
  \item $\rel{R}_D$ and $\rel{R}_a$, for each $a \in \agents$, are
    binary relations on $S$;
  \item $\rel{R}_C$ is the transitive closure of $\rel{R}_D \union
    \bigunion_{a \in \agents} \rel{R}_a$.
  \end{enumerate}
\end{definition}
\begin{definition}
  \label{def:maef}
  A \de{multi-agent epistemic frame} (MAEF, for short) is a MAES\
  $\kframe{F} = (\agents, S, \set{\rel{R}_a}_{a \in \agents},
  \rel{R}_D, \rel{R}_C)$, where
  \begin{description}
    \itemsep-4pt
  \item[(a)] $\rel{R}_D$ and $\rel{R}_a$, for every $a
    \in \agents$, are equivalence relations on $S$;
  \item[(b)] $\rel{R}_D = \biginter_{a \in \agents} \rel{R}_a$.
  \end{description}
  If condition (b) above is replaced with
  \begin{description}
  \item[($b'$)] $\rel{R}_D \subseteq \biginter_{a \in \agents} \rel{R}_a$,
  \end{description}
  then \kframe{F} is a \de{multi-agent epistemic pseudo-frame}.
\end{definition}
Notice that in (pseudo-)frames condition 4 of
definition~\ref{def:maes} is equivalent to the requirement that
$\rel{R}_C$ is the transitive closure of $\bigunion_{a \in \agents}
\rel{R}_a$. Also notice that, as in any MAEF each $\rel{R}_a$ is an
equivalence relation, $\rel{R}_C$ is also an equivalence relation.
\begin{definition}
  \label{def:maem}
  A \de{multi-agent epistemic model} (MAEM, for short) is a tuple
  $\mmodel{M} = (\kframe{F}, \ap, L)$, where
  \begin{description}
    \itemsep-4pt
  \item[($i$)] $\kframe{F}$ is a MAEF;
  \item[($ii$)] $\ap$ is a (possibly, infinite) set of atomic propositions;
  \item[($iii$)] $L: S \mapsto \power{\ap}$, is a \de{labeling
      function}, where $L(s)$ is the set of all atomic propositions
    that are declared true at $s$.
  \end{description}
  If condition (i) above is replaced by the requirement that
  $\kframe{F}$ is a multi-agent epistemic pseudo-frame, then
  $\mmodel{M}$ is a \de{multi-agent epistemic pseudo-model}
  (pseudo-MAEM).
\end{definition}


The satisfaction relation between (pseudo-)MAEMs and formulae is
defined in the standard way.  In particular,
\begin{itemize}
  \itemsep-4pt
\item \sat{M}{s}{\knows{a} \vp} iff $(s, t) \in \rel{R}_a$ implies
  \sat{M}{t}{\vp};
\item \sat{M}{s}{\distrib{} \vp} iff $(s, t) \in \rel{R}_D$ implies
  \sat{M}{t}{\vp};
\item \sat{M}{s}{\common{} \vp} iff $(s, t) \in \rel{R}_C$ implies
  \sat{M}{t}{\vp}.
\end{itemize}
The truth condition for the operator $\common{}$ can be paraphrased in
terms of reachability.  Let $\kframe{F}$ be a (pseudo-)frame with
state space $S$ and let $s, t \in S$.  We say that \emph{$t$ is
  reachable from $s$} if there exists a sequence $s = s_0 , s_1,
\ldots, s_{n-1}, s_n = t$ of elements of $S$ such that, for every $0
\leq i < n$, there exists $a \in \agents$ such that $(s_i, s_{i+1})
\in R_{a}$.  It is then easy to see that the following truth condition
for $\common{}$ is equivalent in (pseudo-)MAEMs to the one given
above:
\begin{itemize}
\item \sat{M}{s}{\common{} \vp} iff \sat{M}{t}{\vp} whenever $t$ is
  reachable from $s$.
\end{itemize}

Notice that if $\agents = \set{a}$, then the formulae $\knows{a} \vp
\equivalence \distrib{} \vp$ and $\knows{a} \vp \equivalence \common{}
\vp$ are valid for all $\vp \in \lang$, so the one-agent case is
trivialized.  Thus, we assume throughout the remainder of the paper
that $\agents$ contains at least 2 agents.

\begin{definition}[Satisfiability and validity]
  \mbox{}
  \begin{itemize}
    \itemsep-4pt
  \item Let $\vp \in \lang$ and $\mmodel{M}$ be a MAEM.  We say that
    $\vp$ is \de{satisfiable} in $\mmodel{M}$ if \sat{M}{s}{\vp} holds
    for some $s \in \mmodel{M}$ and that $\vp$ is \de{valid} in
    $\mmodel{M}$ if \sat{M}{s}{\vp} holds for every $s \in
    \mmodel{M}$.
  \item Let $\vp \in \lang$ and \cf{M} be a class of models. We say
    that $\vp$ is \de{satisfiable} in \cf{M} if \sat{M}{s}{\vp} holds
    for some $\mmodel{M} \in \cf{M}$ and some $s \in \mmodel{M}$ and
    that $\vp$ is \de{valid} in \cf{M} if \sat{M}{s}{\vp} holds for
    every $\mmodel{M} \in \cf{M}$ and every $s \in \mmodel{M}$.
  \end{itemize}
\end{definition}

The goal of this paper is to develop a sound, complete, and
complexity-optimal tableau-based decision procedure for testing
satisfiability, and hence also validity, of formulas of $\lang$ in the
class of all MAEMs; in other words, the procedure tests for the
belonging of formulae of $\lang$ to the logic \maelcd, which is the
logic of all such models.

\section{Hintikka structures}

The ultimate purpose of the tableau procedure we develop is to check
if the input formula is satisfiable in a MAEM. However, the tableau
attempts not to directly construct a MAEM for the input formula, but
to build a more general kind of semantic structure, viz. a
\emph{Hintikka structure} (which are, therefore, used in proving
completeness of our tableaux). The basic difference between models and
Hintikka structures is that while models determine the truth of every
formula of the language at every state, Hintikka structures only
provide truth values of the formulae relevant to the evaluation of a
fixed formula $\theta$.  Another important difference is that the
accessibility relations in models must satisfy the explicitly stated
conditions of definition~\ref{def:maef}, while in Hintikka structures
we only impose conditions on the sets of formulas in the labels of the
states, which correspond to the desirable conditions on the
accessibility relations. Even though no conditions are implicitly
imposed on the accessibility relations themselves, the labeling is
done is such a way that every Hintikka structure generates, by a
construction described in the proof of
lemma~\ref{lm:hintikka_to_models}, a MAEM in such a way that the
``truth'' of the formulas in the labels is preserved in the resultant
model (whose relations satisfy all conditions of
definition~\ref{def:maef}).

To define Hintikka structures, we need the following auxiliary notion,
inspired by~\cite{HM92}.

\begin{definition}
  \label{def:fully_expanded}
  A set $\D \subseteq \lang$ is \de{fully expanded} if it satisfies
  the following conditions ($\subf{\vp}$ stands for the set of
  subformulae of the formula $\vp$):
  \begin{itemize}
    \itemsep-3pt
  \item if $\neg \neg \vp \in \D$, then $\vp \in \D$;
  \item if $\vp \con \psi \in \D$, then $\vp \in \D$ and $\psi \in
    \D$;
  \item if $\neg (\vp \con \psi) \in \D$, then $\neg \vp \in \D$ or
    $\neg \psi \in \D$;
  \item if $\knows{a} \vp \in \D$, for some $a \in \agents$, then
    $\distrib{} \vp \in \D$;
  \item if $\distrib{} \vp \in \D$, then $\vp \in \D$;
  \item if $\common{} \vp \in \D$, then $\knows{a} (\vp \con \common{}
    \vp) \in \D$ for every $a \in \agents$;
  \item if $\neg \common{} \vp \in \D$, then $\neg \knows{a} (\vp \con
    \common{} \vp) \in \D$ for some $a \in \agents$;
  \item if $\vp \in \D$ and $\psi \in \subf{\vp}$ is of the form
    $\knows{a} \chi$ or $\distrib{} \chi$, then either $\psi \in \D$
    or $\neg \psi \in \D$.
  \end{itemize}
\end{definition}

\begin{definition}
  A \de{multi-agent epistemic Hintikka structure} (MAEHS for short)
  is a tuple $(\agents, S, \set{\rel{R}_a}_{a \in \agents}, \rel{R}_D,
  \rel{R}_C, H)$ such that

  \begin{itemize}
  \item $(\agents, S, \set{\rel{R}_a}_{a \in \agents}, \rel{R}_D,
    \rel{R}_C)$ is a MAES;
  \item $H$ is a labeling of the elements of $S$ with formulae of
    \lang\ that satisfies the following constraints:
    \begin{description}
      \itemsep-2pt
    \item[H1] if $\neg \vp \in H(s)$, then $\vp \notin H(s)$;
    \item[H2] $H(s)$ is fully expanded, for every $s \in S$;
    \item[H3] if $\knows{a} \vp \in H(s)$ and $(s, t) \in \rel{R}_a$,
      then $\vp \in H(t)$;
    \item[H4] if $\neg \knows{a} \vp \in H(s)$, then there exists $t
      \in S$ such that $(s, t) \in \rel{R}_a$ and $\neg \vp \in H(t)$;
    \item[H5] if $(s, t) \in \rel{R}_a$, then $\knows{a} \vp \in H(s)$
      iff $\knows{a} \vp \in H(t)$;
    \item[H6] if $\distrib{} \vp \in H(s)$ and $(s, t) \in \rel{R}_D$,
      then $\vp \in H(t)$;
    \item[H7] if $\neg \distrib{} \vp \in H(s)$, then there exists $t
      \in S$ such that $(s, t) \in \rel{R}_D$ and $\neg \vp \in H(t)$;
    \item[H8] if $(s, t) \in \rel{R}_D$, then $\distrib{} \vp \in
      H(s)$ iff $\distrib{} \vp \in H(t)$, and $\knows{a} \vp \in
      H(s)$ iff $\knows{a} \vp \in H(t)$, for every $a \in \agents$;
    \item[H9] if $\neg \common{} \vp \in H(s)$, then there exists $t
      \in S$ such that $(s, t) \in \rel{R}_C$ and $\neg \vp \in H(t)$.
    \end{description}
  \end{itemize}
\end{definition}


\begin{definition}
  Let $\theta \in \lang$ and $\hintikka{H}$ be a MAEHS with state
  space $S$.  We say that \hintikka{H} is a MAEHS for $\theta$ if
  $\theta \in H(s)$ for some $s \in S$.
\end{definition}

Now we will prove that $\theta \in \lang$ is satisfiable in the class of
all MAEMs iff there exists a MAEHS for $\theta$.  This will allow us to
design our tableau procedure to test for the existence of a
MAEHS, rather than a MAEM, for the input formula.

Given a MAEM $\mmodel{M}$ with a labeling function $L$, we define the
\emph{extended labeling function} $L^+: S \mapsto \power{\lang}$ on
\mmodel{M} as follows: $L^+(s) = \crh{\vp}{\sat{M}{s}{\vp}}$.  Then,
the following is straightforward.

\begin{lemma}
  \label{lm:models_to_hintikka}
  Let $\mmodel{M} = (\agents, S, \set{\rel{R}_a}_{a \in \agents},
  \rel{R}_D, \rel{R}_C, L)$ be a MAEM satisfying $\theta$ and let
  $L^+$ be an extended labeling on \mmodel{M}.  Then, $(\agents, S,
  \set{\rel{R}_a}_{a \in \agents}, \rel{R}_D, \rel{R}_C, L^+)$ is a
  MAEHS for $\theta$.
\end{lemma}

Next, we prove the opposite direction.

\begin{lemma}
  \label{lm:hintikka_to_models}
  Let $\theta \in \lang$ be such that there exists a MAEHS for
  $\theta$.  Then, $\theta$ satisfiable in a MAEM.
\end{lemma}

\begin{proof}
  Let $\theta \in \lang$ and $\hintikka{H} = (\agents, S,
  \set{\rel{R}_a}_{a \in \agents}, \rel{R}_D, \rel{R}_C, H)$ be an
  MAEHS for $\theta$.  First, we define, using \hintikka{H}, a
  pseudo-MAEM \mmodel{M'} satisfying $\theta$; then, we turn
  \mmodel{M'} into a MAEM satisfying $\theta$.

  \mmodel{M'} is defined as follows. First, for every $a \in \agents$,
  let $\rel{R}'_a$ be the reflexive, symmetric, and transitive closure
  of $\rel{R}_a \union \rel{R}_D$; let $\rel{R}'_D$ be the reflexive,
  symmetric, and transitive closure of $\rel{R}_D$; and let
  $\rel{R}'_C$ be the transitive closure of $\bigunion_{a \in \agents}
  \rel{R}'_a$. (Notice that $\rel{R}_C \subseteq \rel{R}'_C$.) Second,
  let $\ap = \crh{p \in H(t)}{t \in S \text{ and } p \text{ is an
      atomic proposition}}$. Finally, let $L(s) = H(s) \inter \ap$ for
  every $s \in S$.  It is then straightforward to check that
  $\mmodel{M'} = (\agents, S, \set{\rel{R}'_a}_{a \in \agents},
  \rel{R}'_D, \rel{R}'_C, \ap, L)$ is a pseudo-MAEM (recall
  definition~\ref{def:maem}).

  Next, we prove, by induction on the structure of $\chi \in \lang$
  that, for every $s \in S$ and every $\chi \in \lang$, the following
  hold:

  i) $\chi \in H(s) \text{ implies } \sat{M'}{s}{\chi}$, and

  ii) $\neg \chi \in H(s) \text{ implies } \sat{M'}{s}{\neg \chi}$.

  Let $\chi$ be some $p \in \ap$.  Then, $p \in H(s)$ implies $p \in
  L(s)$ and, thus, \sat{M'}{s}{p}; if, on the other hand, $\neg p \in
  H(s)$, then due to (H1), $p \notin H(s)$ and thus $p \notin L(s)$;
  hence, \sat{M'}{s}{\neg p}.

  Assume that the claim holds for all subformulae of $\chi$; then, we have
  to prove that it holds for $\chi$, as well.

  Suppose that $\chi$ is $\neg \vp$.  If $\neg \vp \in H(s)$, then the
  inductive hypothesis immediately gives us $\sat{M'}{s}{\neg \vp}$;
  if, on the other hand, $\neg \neg \vp \in H(s)$, then by virtue of
  (H2), $\vp \in H(s)$ and hence, by inductive hypothesis,
  $\sat{M'}{s}{\vp}$ and thus $\sat{M'}{s}{\neg \neg \vp}$.

  The case of $\chi = \vp \con \psi$ is straightforward, using (H2).

  Suppose that $\chi$ is $\knows{a} \vp$.  Assume, first, that
  $\knows{a} \vp \in H(s)$.  In view of inductive hypothesis, it
  suffices to show that $(s, t) \in \rel{R}'_a$ implies $\vp \in H(t)$.
  So, assume that $(s, t) \in \rel{R}'_a$.  There are two cases to
  consider.  If $s = t$, then the conclusion immediately follows from
  (H2).  If, on the other hand, $s \ne t$, then there exists an
  undirected path from $s$ to $t$ along the relations $\rel{R}_a$ and
  $\rel{R}_D$.  Then, in view of (H5) and (H8), $\knows{a} \vp \in
  H(t)$; hence, by (H2), $\vp \in H(t)$.

  Assume, next, that $\neg \knows{a} \vp \in H(s)$.  In view of the
  inductive hypothesis, it suffices to show that there exist $t \in
  S$ such that $(s, t) \in \rel{R}'_a$ and $\neg \vp \in H(t)$.  By
  (H4), there exists $t \in S$ such that $(s, t) \in \rel{R}_a$ and
  $\neg \vp \in H(t)$.  As $\rel{R}_a \subseteq \rel{R}'_a$, the
  desired conclusion follows.

  The case of $\chi = \distrib{} \vp$ is very similar to the previous
  one and is left to the reader.

  Suppose now that $\chi$ is $\common{} \vp$.  Assume that $\common{}
  \vp \in H(s)$.  In view of the inductive hypothesis, it suffices to
  show that if $(s, t) \in \rel{R}'_C$, then $\vp \in H(t)$. So,
  assume that $(s, t) \in \rel{R}'_C$, i.e., either $s = t$ or, for
  some $n \geq 1$, there exists a sequence of states $s = s_0, s_1,
  \ldots, s_{n-1}, s_n = t$ such that, for every $0 \leq i < n$,
  either there exists $a \in \agents$ such that $(s_i, s_{i+1}) \in
  \rel{R}_a$ or $(s_i, s_{i+1}) \in \rel{R}_D$. In the former case,
  the desired conclusion follows from (H2); in the latter, it follows
  from (H2), (H3), and (H8).

  Assume, on the other hand, that $\neg \common{} \vp \in H(s)$.
  Then, the desired conclusion follows from (H9), the fact that
  $\rel{R}_C \subseteq \rel{R}'_C$, and inductive hypothesis.

  To finish the proof of the lemma, we convert $\mmodel{M}'$ into a
  MAEM $\mmodel{M}''$ in a truth-preserving way.  To that end, we use
  a variation of the construction known as tree-unwinding (see, for
  example, \cite{GorOtto07}; first applied in the context of epistemic
  logics with the operator of distributed knowledge in~\cite{FHV92}
  and \cite{vdHM92}).  The only difference between our construction
  and the standard tree-unwinding is that, in the tree we produce, all
  edges labeled by $D$ (representing the tree's relation
  $\rel{R}^{T}_D$) also get labeled (unlike in the standard
  tree-unwinding) by \emph{all} agents in $\agents$, too;
  all other transitions are labeled by single agents, as in the
  standard tree-unwinding.  To obtain $\mmodel{M}''$, we take
  $\rel{R}''_D$ to be the reflexive, symmetric, and transitive closure
  of $\rel{R}^{T}_D$ and $\rel{R}''_a$, for every $a \in \agents$, to
  be the reflexive, symmetric, and transitive closure of
  $\rel{R}^{T}_a$; finally, we take $\rel{R}''_C$ to be the reflexive
  closure of $\bigunion_{a \in \agents} \rel{R}''_a$.  It is routine
  to check that $\mmodel{M}''$ is bisimilar to $\mmodel{M}'$ and,
  therefore, satisfies $\theta$ at its root.  To complete the proof,
  all we have to show is that $\mmodel{M}''$ is a MAEM; i.e., the
  equality $\rel{R}''_D = \biginter_{a \in \agents} \rel{R}''_a$
  holds.  The left-to-right direction is immediate from the
  construction.  For the right-to-left direction assume that $(s, t)
  \in \rel{R}''_a$ holds for every $a \in \agents$; i.e, there is an
  undirected path between $s$ and $t$ along $\rel{R}^T_a$ for every $a
  \in \agents$.  As we are in a tree and $\agents$ contains at least
  two agents, this is only possible if there is an undirected path
  between $s$ and $t$ along $\rel{R}^T_D$ since we only connected
  nodes of the tree by multiple agent relations if these nodes were
  connected by $\rel{R}^T_D$.  Therefore, $(s, t) \in \rel{R}''_D$, as
  desired.
\end{proof}

\begin{theorem}
  \label{thr:sat_equal_hintikka}
  Let $\theta \in \lang$.  Then, $\theta$ is satisfiable in a MAEM
  iff there exists a MAEHS for $\theta$.
\end{theorem}

\begin{proof}
  Immediate from
  lemma~\ref{lm:models_to_hintikka} and
  lemma~\ref{lm:hintikka_to_models}.
\end{proof}

\section{Tableau procedure for \maelcd}

Traditionally, tableaux work by decomposing the formula whose
satisfiability is being tested into ``semantically simpler''
formulae. In the classical propositional case, ``semantically
simpler'' implies ``smaller'', which by itself guarantees termination
of the procedure. Another feature of the tableau method for the
classical propositional logic is that this decomposition into simpler
formulae results in a simple tree, representing an exhaustive search
for a model---or, to be more precise, a Hintikka set (the classical
analogue of Hintikka structures)---for the input formula.  If at least
one leaf of the tree produces a Hintikka set for the input formula,
the search has succeeded and the formula is pronounced satisfiable.

These two defining features of the classical tableau method do not
emerge unscathed when the method is applied to logics containing fixed
point operators, such as $\common{}$ (or, for example, the $\until$
and $\neg \Box$ operators of the linear-time temporal logic
\LTL). Firstly, decomposing (in accordance with the clauses in the
definition of a fully expanded set above) of formulae of the form
$\common{} \vp$ produces formulae of the form $\knows{a} (\vp \con
\common{} \vp)$, which are ``semantically simpler'', but not smaller
than the original formula.  Hence, we cannot take termination for
granted and need to take special precautions to guarantee it---in our
tableaux, we do so by deploying prestates, whose role is to ensure
that the whole construction is finite.  Secondly, in the classical
case, the only reason why it might turn out to be impossible to
produce a Hintikka set for the input formula is that every attempt to
build such a set results in a collection of formulae containing an
inconsistency.  In the case of \maelcd, there are other such reasons;
the most important of them has to do with eventualities: semantically,
the truth of an eventuality $\neg \common{} \vp$ at state $s$ of a
model requires that there is a path form $s$ to a state $t$ satisfying
$\neg \vp$.  The analogue of this semantic condition in the tableau we
refer to as \emph{realization of eventualities}.  Apart from
consistency requirement on a ``good'' tableau, all eventualities in
such a tableau should be realized.  (A third, more technical reason
why a tableau might fail to represent a MAEHS will be mentioned in due
course.)

\subsection{Overview of the tableau procedure}

In essence, the tableau procedure for testing a formula $\theta \in
\lang$ for satisfiability is an attempt to construct a non-empty graph
$\tableau{T}^{\theta}$, called a \fm{tableau}, representing all
possible MAEHSs for $\theta$ (in the sense made precise later on).  If
the attempt is successful, $\theta$ is pronounced satisfiable;
otherwise, it is declared unsatisfiable.

The tableau procedure consists of three major phases: \fm{construction
  phase}, \fm{prestate elimination phase}, and \fm{state elimination
  phase}.  Accordingly, we have three types of tableau rules:
construction rules, a prestate elimination rule, and state elimination
rules.  The procedure itself essentially specifies in what order and
under what circumstances these rules should be applied.

During the construction phase, the construction rules are used to
produce a directed graph $\tableau{P}^{\theta}$--- called the
\emph{pretableau} for $\theta$---whose set of nodes properly contains
the set of nodes of the tableau $\tableau{T}^{\theta}$ that we are
building.  Nodes of $\tableau{P}^{\theta}$ are sets of formulae, some
of which, called \fm{states}, are meant to represent states of a
Hintikka structure, while others, called \fm{prestates}, fulfill a
purely technical role of to keeping $\tableau{P}^{\theta}$ finite.
During the prestate elimination phase, we create a smaller graph
$\tableau{T}_0^{\theta}$ out of $\tableau{P}^{\theta}$, called the
\fm{initial tableau for $\theta$}, by eliminating all prestates of
$\tableau{P}^{\theta}$ (and tweaking with its edges) since prestates
have already fulfilled their function: as we are not going to add any
more nodes to the graph built so far, the possibility of producing an
infinite structure is no longer a concern.  Lastly, during the state
elimination phase, we remove from $\tableau{T}_0^{\theta}$ all states,
if any, that cannot be satisfied in any MAEHS, for one of the
following three reasons: either the state is inconsistent, or it
contains an unrealized eventuality, or it does not have all successors
needed for its satisfaction.  The elimination procedure results in a
(possibly empty) subgraph $\tableau{T}^{\theta}$ of
$\tableau{T}_0^{\theta}$, called the \de{final tableau for
  $\theta$}. Then, if we have some state $\Delta$ in
$\tableau{T}^{\theta}$ containing $\theta$, we declare $\theta$
satisfiable; otherwise, we declare it unsatisfiable.

\subsection{Construction phase}

At this phase, we build the pretableau $\tableau{P}^{\theta}$ --- a
directed graph whose nodes are sets of formulae, coming in two
varieties: \fm{states} and \fm{prestates}.  States are meant to
represent states of a MAEHS which the tableau attempts to construct,
while prestates are ``embryo states'', which will in the course of the
construction be ``unwound'' into states.  Technically, states are
fully expanded (recall definition~\ref{def:fully_expanded}), while
prestates do not have to be so.

Moreover, $\tableau{P}^{\theta}$ will contain two types of edges.  As
we have already mentioned, our tableaux attempt to produce a MAEHS for
the input formula; in this attempt, they set in motion an exhaustive
search for such a MAEHS. One type of edge, depicted by unmarked double
arrows $\brancharrow$, will represent this exhaustive search dimension
of our tableaux.  Exhaustive search looks for all possible
alternatives, and in our tableaux the alternatives will arise when we
unwind prestates into states; thus, when we draw an unmarked arrow
from a prestate \G\ to states $\D$ and $\D'$ (depicted as $\G
\brancharrow \D$ and $\G \brancharrow \D'$, respectively), this
intuitively means that, in any MAEHS, a state satisfying \G\ has to
satisfy at least one of $\D$ and $\D'$.

Given a set $\G \subseteq \lang$, we say that $\D$ is a \de{minimal
  fully expanded extension of \G} if $\D$ is fully expanded, $\G
\subseteq \D$, and no $\D'$ is such that $\G \subseteq \D' \subset \D$
and $\D'$ is fully expanded.

Our first construction rule, \Rule{SR}, tells us how to create states
from prestates. (Throughout the presentation of the rules, the reader
can refer to the example given below to see how they are applied in
particular cases.)

\medskip

\Rule{SR} Given a prestate $\G$, do the following:

\begin{enumerate}
  \itemsep-4pt
\item add to the pretableau all minimal fully expanded extensions
  $\D$ of $\G$ as \de{states};

\item for each so obtained state $\D$, put $\G \brancharrow \D$;

\item if, however, the pretableau already contains a state $\D'$ that
  coincides with $\D$, do not create another copy of $\D'$, but only
  put $\G \brancharrow \D'$.
\end{enumerate}

We denote the finite set of states created by applying \Rule{SR} to a
prestate $\G$ by $\st{\G}$.

The second type of edge featuring in our tableaux represents
accessibility relations in MAEHSs.  Accordingly, this type of edge
will be represented by single arrows marked with formulas whose
presence in the source state requires the existence of a target state
reachable by a particular relation.  As there are two such kinds of
formulae, $\neg \knows{a} \vp$ and $\neg \distrib{} \vp$ (see
conditions (H4) and (H7) in the definition of MAEHS), we will have
single arrows marked by formulas of one of these two types.
Intuitively if, say $\neg \knows{a} \vp \in \D$, then we need some
prestate $\G$ containing $\neg \vp$ to be accessible by a relation
$\rel{R}_a$; however, we mark this single arrow not just by agent $a$,
but by formula $\neg \knows{a} \vp$, which helps us remember not just
what relation connects states satisfying $\D$ and $\G$, but why we had
to create this particular $\G$.  This information will prove crucial
when we start eliminating prestates and then states.

The two remaining construction rules, \Rule{KR} and \Rule{DR}, tell us
how to create prestates from states.  These rules do not apply to
patently inconsistent states as such states can not be satisfied in
any MAEHS.

\medskip

\Rule{KR} Given a state $\D$ such that $\neg \knows{a}
\vp \in \D$, for some $a \in \agents$, and there is no $\chi \in \lang$ such that both $\chi \in \D$ and $\neg \chi \in \D$,  do the following:

\begin{enumerate}
  \itemsep-4pt
\item create a new prestate $\G = \set{\neg \vp} \union \crh{\knows{a}
    \psi}{\knows{a} \psi \in \D} \union \crh{\neg \knows{a} \psi}{\neg
    \knows{a} \psi \in \D}$;
\item connect $\D$ to $\G$ with $\stackrel{\neg \knows{a}
    \vp}{\longrightarrow}$;
\item if, however, the tableau already contains a prestate $\G' = \G$,
  do not add to it another copy of $\G'$, but simply connect $\D$ to
  $\G'$ with $\stackrel{\neg \knows{a} \vp}{\longrightarrow}$.
\end{enumerate}

\Rule{DR} Given a state $\D$ such that $\neg \distrib{}
\vp \in \D$ and there is no $\chi \in \lang$ such that both $\chi \in \D$ and $\neg \chi \in \D$, do the following:

\begin{enumerate}
  \itemsep-4pt
\item create a new prestate $\G = \set{\neg \vp} \union
  \crh{\distrib{} \psi}{\distrib{} \psi \in \D} \union \crh{\neg
    \distrib{} \psi}{\neg \distrib{} \psi \in \D} \union
  \crh{\knows{a} \chi}{\knows{a} \chi \in \D, a \in \agents} \union \crh{\neg \knows{a} \chi}{\neg \knows{a} \chi \in \D, a \in \agents}$;
\item connect $\D$ to $\G$ with $\stackrel{\neg \distrib{}
    \vp}{\longrightarrow}$;
\item if, however, the tableau already contains a prestate $\G' = \G$,
  do not add to it another copy of $\G'$, but simply connect $\D$ to
  $\G'$ with $\stackrel{\neg \distrib{} \vp}{\longrightarrow}$.
\end{enumerate}

It should be noted that, in the pretableau, we never create in one go
full-fledged successors for states; i.e., we never draw a marked arrow
from state to state; such arrows always go from states to prestates.
On the other hand, unmarked arrows connect prestates to states.

When building a tableau for a formula $\theta$, the construction stage
starts off with creating a single prestate \set{\theta}.  Afterwards,
we alternate between applying rules creating states and those creating
prestates: first, \Rule{SR} is applied to the prestates created at the
previous stage of the construction, then \Rule{KR} and \Rule{DR} are
applied to the states created at the previous stage.  The construction
phase comes to an end when every prestate required to be added to the
pretableau has already been added (as prescribed in point 3 of
\Rule{SR}), or when we end up with states to which neither \Rule{KR}
nor \Rule{DR} is applicable (i.e. states not containing formulas of
the form $\neg \knows{a} \vp$ or $\neg \distrib{} \vp$ or containing
patent inconsistencies).

\subsection{Termination of construction phase}

As we identify states and prestates whenever possible, to prove that
the above procedure terminates, it suffices to establish that there
are only finitely many possible states and prestates.  To that end we
use the concept of the extended closure of a formula $\theta$.

\begin{definition}
  Let $\theta \in \lang$.  The \de{closure} of $\theta$, denoted
  $\cl{\theta}$, is the least set of formulae such that:
  \begin{itemize}
    \itemsep-4pt
  \item $\theta \in \cl{\theta}$;
  \item $\cl{\theta}$ is closed under subformulae;
  \item if $\knows{a} \vp \in \cl{\theta}$ for some $a \in \agents$,
    then $\distrib{} \vp \in \cl{\theta}$;
  \item if $\common{} \vp \in \cl{\theta}$, then $\knows{a} (\vp \con
    \common{} \vp) \in \cl{\theta}$ for every $a \in \agents$.
  \end{itemize}
\end{definition}

\begin{definition}
  \label{def:extended_closure}
  Let $\theta \in \lang$.  The \de{extended closure} of $\theta$,
  denoted $\ecl{\theta}$, is the least set such that if $\vp \in
  \cl{\theta}$, then $\vp, \neg \vp \in \ecl{\theta}$.

\end{definition}

It is straightforward to check that $\ecl{\theta}$ if finite for every
$\theta$ and that all state and prestates of $\tableau{P}^{\theta}$
are subsets of $\ecl{\theta}$; hence, their number is finite.

\subsection{Prestate elimination phase}

At this phase of the tableau procedure, we remove from
$\tableau{P}^{\theta}$ all prestates and all unmarked arrows,
by applying the following rule:

\medskip

\Rule{PR} For every prestate $\G$ in $\tableau{P}^{\theta}$, do the
following:

\begin{enumerate}
  \itemsep-4pt
\item remove $\G$ from $\tableau{P}^{\theta}$;
\item if there is a state $\D$ in $\tableau{P}^{\theta}$ with $\D
  \stackrel{\chi}{\longrightarrow} \G$, then for every state $\D' \in
  \st{\G}$, put $\D \stackrel{\chi}{\longrightarrow} \D'$;
\end{enumerate}

We call the graph obtained by applying \Rule{PR} to
$\tableau{P}^{\theta}$ the \emph{initial tableau}, denoted by
$\tableau{T}_0^{\theta}$.

\subsection{State elimination phase}

During this phase, we remove from $\tableau{T}_0^{\theta}$ nodes that
cannot be satisfied in any MAEHS.  There are three reasons why a
state $\D$ of $\tableau{T}_0^{\theta}$ can turn out to be
unsatisfiable: $\D$ contains an inconsistency, \emph{or} satisfiability
of $\D$ requires satisfiability of some other unsatisfiable
``successor'' states, \emph{or} $\D$ contains an eventuality that is
not realized in the tableau. Accordingly, we have three elimination rules,
\Rule{E1}--\Rule{E3}.

Technically, the state elimination phase is divided into stages; at
stage $n+1$ we remove from the tableau $\tableau{T}_n^{\theta}$
obtained at the previous stage exactly one state, by applying one of
the elimination rules, thus obtaining the tableau
$\tableau{T}_{n+1}^{\theta}$. We now state the rules governing the
process.  The set of states of the tableau $\tableau{T}_{m}^{\theta}$ is
denoted by $S_m^{\theta}$.

\bigskip

\Rule{E1} If $\set{\vp, \neg \vp} \subseteq \D \in S_n^{\theta}$, then
obtain $\tableau{T}_{n+1}^{\theta}$ by eliminating $\D$ from
$\tableau{T}_n^{\theta}$.

\bigskip

\Rule{E2} If $\D$ contains a formula $\chi$ of the form $\neg
\knows{a} \vp$ or $\neg \distrib{} \vp$ and all states reachable
from $\D$ by single arrows marked by $\chi$ have been eliminitated at
previous stages, obtain $\tableau{T}_{n+1}^{\theta}$ by eliminating
$\D$ from $\tableau{T}_n^{\theta}$.

\bigskip

To formulate the third elimination rule, we need the concept of
eventuality realization.  We say that $\neg \common{} \vp$ is realized
at $\D$ in $\tableau{T}^{\theta}_n$ if there exists a path $\D = \D_0,
\D_1, \ldots, \D_m$ such that $\neg \vp \in \D_m$ and, for every $0
\leq i < m$, there exist $\chi$ such that $\D_i
\stackrel{\chi}{\longrightarrow} \D_{i+1}$.

Realization of eventuality $\neg \common{} \vp$ at $\D$ in
$\tableau{T}^{\theta}_n$ can be easily checked by computing the
\emph{rank} of every $\D \in S^{\theta}_n$ with respect to $\neg
\common{} \vp$ in $\tableau{T}^{\theta}_n$, denoted by
$\mathbf{rank}(\D, \neg \common{} \vp, \tableau{T}_n^{\theta})$.
Intuitively, the rank of $\D$ in $\tableau{T}_n^{\theta}$ represents
the length of the longest path in $\tableau{T}_n^{\theta}$ from $\D$
to a state containing $\neg \vp$.  If no such path exists, the rank of
$\D$ is $\omega$ (the first infinite ordinal).  Formally, the rank is
computed as follows.  At first, if $\neg \vp \in \D$, set
$\mathbf{rank}(\D, \neg \common{} \vp, \tableau{T}_n^{\theta}) = 0$;
otherwise, set $\mathbf{rank}(\D, \neg \common{} \vp,
\tableau{T}_n^{\theta}) = \omega$.  Afterwards, repeat the following
procedure until no changes in the rank of any state occurs:
$\mathbf{rank}(\D, \neg \common{} \vp, \tableau{T}_n^{\theta}) = 1 +
\max \set{r_{\chi}}$, where $r_{\chi} = \min \crh{\mathbf{rank}(\D',
  \neg \common{} \vp, \tableau{T}_n^{\theta})}{\D
  \stackrel{\chi}{\longrightarrow} \D'}$.  Now, we can state our last
rule.

\medskip

\Rule{E3} If $\D \in S_n^{\theta}$ contains an eventuality
  $\neg \common{} \vp$ that is not realized at $\D$ in
  $\tableau{T}_n^{\theta}$ (i.e., if $\mathbf{rank}(\D, \neg \common{}
  \vp, \tableau{T}_n^{\theta}) = \omega$), then obtain
  $\tableau{T}_{n+1}^{\theta}$ by removing $\D$ from
  $\tableau{T}_n^{\theta}$.

\medskip

We have thus far described the individual rules; to describe the state
elimination phase as a whole, it is crucial to specify the order of
their application.

First, we apply \Rule{E1} to all states of
$\tableau{T}_0^{\theta}$; it is clear that, once this is done, we do not
need to go back to \Rule{E1} again.  The cases of \Rule{E2} and
\Rule{E3} are slightly more involved.  Having applied \Rule{E3} to the
states of the tableau, we could have removed, for some $\D$, all
states accessible from it along the arrows marked with some formula
$\chi$; hence, we need to reapply \Rule{E2} to the resultant tableau
to get rid of such $\D$'s.  Conversely, having applied \Rule{E2}, we
could have removed some states that were instrumental in realizing
certain eventualities; hence, having applied \Rule{E2}, we need to
reapply \Rule{E3}.  Furthermore, we can't stop the procedure unless we
have checked that \emph{all} eventualities are realized.  Thus, what
we need is to apply \Rule{E3} and \Rule{E2} in a dovetailed sequence
that cycles through all eventualities. More precisely, we arrange
all eventualities occurring in the tableau obtained from
$\tableau{T}_0^{\theta}$ after having applied \Rule{E1} in the list
$\xi_1, \ldots, \xi_m$.  Then, we proceed in cycles. Each cycle
consists of alternatingly applying \Rule{E3} to the pending
eventuality, and then applying \Rule{E2} to the tableau resulting from
that application, until all eventualities have been dealt with;
once we reach $\xi_m$, we loop back to $\xi_1$.  The cycles are
repeated until, having gone through the whole cycle, we have not
removed any states.

Once that happens, the state elimination phase is over. We call the resultant
graph the \fm{final tableau for $\theta$} and denote it by
$\tableau{T}^{\theta}$ and its set of states by $S^{\theta}$.
\begin{definition}
  The final tableau $\tableau{T}^{\theta}$ is \de{open} if $\theta \in
  \D$ for some $\D \in S^{\theta}$; otherwise, $\tableau{T}^{\theta}$
  is \de{closed}.
\end{definition}
The tableau procedure returns ``no'' if the final tableau is closed;
otherwise, it returns ``yes'' and, moreover, provides sufficient
information for producing a finite model satisfying $\theta$; that
construction is described in section \ref{sec:completeness}.
\begin{example}
  Let's assume that $\agents = \set{a, b}$ and construct a tableau for
  the formula $\knows{a} p \con \knows{b} p \con \neg \distrib{}
  \common{} p$. The picture below shows the complete pretableau for
  this formula.

\begin{picture}(200,255)(0,245)
   \footnotesize
    \thicklines

    \put(98,490){\makebox(0,0){
        $\G_0$
      }}

    \put(95,485){\line(0,-1){10}}
    \put(96.25,485){\line(0,-1){10}}
    \put(95.75,475){\vector(0,-1){5}}

    \put(98,463){\makebox(0,0){
        $\D_1$
      }}

    \put(95.75,455){\line(0,-1){10}}
    \put(95.75,445){\vector(0,-1){5}}

    \put(105,450){\makebox(0,0){
        {\tiny $\chi_0$ }
      }}

    \put(98,432){\makebox(0,0){
        {$\G_1$ }
      }}

    \put(84,428){\line(-1,-1){17}}
    \put(86.5,429){\line(-1,-1){19}}
    \put(68,411){\vector(-1,-1){5}}

    \put(105,428){\line(1,-1){17}}
    \put(106.5,428){\line(1,-1){17}}
    \put(122,412){\vector(1,-1){5}}

    \put(55,405){\makebox(0,0){
        {$\D_2$ }
      }}

    \put(50,400){\line(-1,-1){20}}
    \put(31,381){\vector(-1,-1){5}}
    \put(36,393){\makebox(0,0){
        {\tiny $\chi_1$ }
      }}

    \put(53,400){\line(1,-1){22}}
    \put(72,381){\vector(1,-1){5}}
    \put(65,396){\makebox(0,0){
        {\tiny $\chi_0$ }
      }}

    \put(28,370){\makebox(0,0){
        {$\G_2$ }
      }}
    \put(18,364){\line(-1,-1){10}}
    \put(19.5,364){\line(-1,-1){10}}
    \put(9.75,355){\vector(-1,-1){5}}
    \put(7,342){\makebox(0,0){
        {$\D_5$ }
      }}
     \qbezier(-2, 342)(-12, 355)(20, 370)
    \put(17,368){\vector(1,1){5}}
    \put(-10,355){\makebox(0,0){
        {\tiny $\chi_1$ }
      }}
    \put(24,364){\line(0,-1){10}}
    \put(25.25,364){\line(0,-1){10}}
    \put(24.75,355){\vector(0,-1){5}}
    \put(27,342){\makebox(0,0){
        {$\D_4$ }
      }}
    \put(29,364){\line(1,-1){10}}
    \put(30.5,364){\line(1,-1){10}}
    \put(38.75,355){\vector(1,-1){5}}
    \put(47,342){\makebox(0,0){
        {$\D_6$ }
      }}
    \qbezier(48, 344)(50, 355)(35, 365)
    \put(38,362){\vector(-1,1){5}}
    \put(45.5,362){\makebox(0,0){
        {\tiny $\chi_1$ }
      }}
    \qbezier(43, 336)(60, 316)(91, 285)
    \put(89,287){\vector(1,-1){5}}
    \put(46,326){\makebox(0,0){
        {\tiny $\chi_2$ }
      }}

    \put(75,370){\makebox(0,0){
        {$\G_3$ }
      }}
    \put(71,364){\line(-1,-1){10}}
    \put(72.5,364){\line(-1,-1){10}}
    \put(62.75,355){\vector(-1,-1){5}}
    \put(63,342){\makebox(0,0){
        {$\D_7$ }
      }}
    \qbezier(58, 345)(50, 355)(67, 366.9)
    \put(64,364.5){\vector(1,1){5}}
   \put(57,364){\makebox(0,0){
       {\tiny $\chi_0$ }
      }}
    \qbezier(56, 343)(41, 360)(46.6, 354)
    \put(64,364.5){\vector(1,1){5}}

    \put(72,364){\line(1,-1){10}}
    \put(73.5,364){\line(1,-1){10}}
    \put(81.75,355){\vector(1,-1){5}}

    \put(85,386){\makebox(0,0){
        {\tiny $\chi_1$ }
      }}
    \put(101,386){\makebox(0,0){
        {\tiny $\chi_2$ }
      }}
    \qbezier(94, 364)(78, 406)(31, 373)
    \put(34.5, 375.5){\vector(-1,-1){5}}
    \qbezier(94, 364)(110, 406)(157, 371)
    \put(154, 374){\vector(1,-1){5}}
    \put(94,349){\line(0,1){15}}
    \qbezier(90, 344)(94, 348)(94, 354)
    \qbezier(98, 344)(94, 348)(94, 354)

    \put(89,342){\makebox(0,0){
        {$\D_8$ }
      }}
    \put(82,338){\vector(1,-1){10}}
    \put(81,332){\makebox(0,0){
        {\tiny $\chi_0$ }
      }}

    \put(138,405){\makebox(0,0){
        {$\D_3$ }
      }}

    \put(135,400){\line(-1,-1){22}}
    \put(115,380){\vector(-1,-1){5}}

    \put(125,397){\makebox(0,0){
        {\tiny $\chi_0$ }
      }}

    \put(138,400){\line(1,-1){22}}
    \put(158,380){\vector(1,-1){5}}

    \put(154,393){\makebox(0,0){
        {\tiny $\chi_2$ }
      }}

    \put(115,370){\makebox(0,0){
        {$\G_4$ }
      }}
    \put(114,364){\line(-1,-1){10}}
    \put(115.5,364){\line(-1,-1){10}}
    \put(105.75,355){\vector(-1,-1){5}}
    \put(106,342){\makebox(0,0){
        {$\D_9$ }
      }}
    \put(107,338){\vector(-1,-1){10}}
    \put(111,332){\makebox(0,0){
        {\tiny $\chi_0$ }
      }}

    \put(116,364){\line(1,-1){10}}
    \put(117.5,364){\line(1,-1){10}}
    \put(125.75,355){\vector(1,-1){5}}
    \put(131,342){\makebox(0,0){
        {$\D_{10}$ }
      }}
    \put(133,364){\makebox(0,0){
        {\tiny $\chi_0$ }
      }}
    \qbezier(130, 344)(138, 351)(123, 365)
    \put(125,363){\vector(-1,1){5}}

    \qbezier(134, 344)(138, 351)(142, 355)
    \put(125,363){\vector(-1,1){5}}

    \put(165,370){\makebox(0,0){
        {$\G_5$ }
      }}

    \put(158,364){\line(-1,-1){10}}
    \put(159.5,364){\line(-1,-1){10}}
    \put(149.75,355){\vector(-1,-1){5}}

    \put(149,342){\makebox(0,0){
        {$\D_{12}$ }
      }}
    \qbezier(144, 347)(135, 356)(155, 366)
    \put(152.5,364){\vector(1,1){5}}
    \put(147,366){\makebox(0,0){
        {\tiny $\chi_2$ }
      }}

    \qbezier(146, 336)(130, 316)(106, 285)
    \put(107.5,287){\vector(-1,-1){5}}
    \put(147,326){\makebox(0,0){
        {\tiny $\chi_1$ }
      }}

    \put(164,364){\line(0,-1){10}}
    \put(165.25,364){\line(0,-1){10}}
    \put(164.75,355){\vector(0,-1){5}}

    \put(167,342){\makebox(0,0){
        {$\D_{11}$ }
      }}

    \put(169,364){\line(1,-1){10}}
    \put(170.5,364){\line(1,-1){10}}
    \put(178.75,355){\vector(1,-1){5}}

    \put(189,342){\makebox(0,0){
        {$\D_{13}$ }
      }}

    \put(183.5,370){\makebox(0,0){
        {\tiny $\chi_2$ }
      }}
    \qbezier(196, 342)(201, 351)(173, 370)
    \put(175,368){\vector(-1,1){5}}

    \put(98,322){\makebox(0,0){
        {$\G_{6}$ }
      }}
    \put(95,318){\line(0,-1){10}}
    \put(96.25,318){\line(0,-1){10}}
    \put(95.75,310){\vector(0,-1){5}}

    \put(100,300){\makebox(0,0){
        {$\D_{14}$ }
      }}
    \put(78,310){\makebox(0,0){
        {\tiny $\chi_0$ }
      }}
    \qbezier(88, 320)(78, 310)(88, 300)
    \put(87,319){\vector(1,1){5}}

    \put(100,280){\makebox(0,0){
        {$\G_7$ }
      }}
    \put(94,274){\line(-1,-1){10}}
    \put(95.5,274){\line(-1,-1){10}}
    \put(85.75,265){\vector(-1,-1){5}}

    \put(97,274){\line(0,-1){10}}
    \put(98,274){\line(0,-1){10}}
    \put(97.5,264){\vector(0,-1){5}}

    \put(100,274){\line(1,-1){10}}
    \put(101.5,274){\line(1,-1){10}}
    \put(110,265){\vector(1,-1){5}}

    \put(82,253){\makebox(0,0){
        {$\D_{16}$ }
      }}
    \put(77,275){\makebox(0,0){
        {\tiny $\chi_1$ }
      }}
    \qbezier(73, 255)(70, 270)(92, 278)
    \put(89,276){\vector(1,1){5}}

    \put(102,253){\makebox(0,0){
        {$\D_{15}$ }
      }}

    \put(122,253){\makebox(0,0){
        {$\D_{17}$ }
      }}
    \put(117,275){\makebox(0,0){
        {\tiny $\chi_2$ }
      }}
    \qbezier(120, 255)(127, 265)(105, 277.5)
    \put(108.5,275){\vector(-1,1){5}}

\end{picture}

{\footnotesize

  $\chi_0 = \neg \distrib{} \common{} p$, $\chi_1 = \neg \knows{a} (p
  \con \common{} p)$, $\chi_2 = \neg \knows{b} (p \con \common{} p)$;

  $\G_0 = \set{\knows{a} p \con \knows{b} p \con \neg \distrib{}
    \common{} p}$;

  $\D_1 = \set{\knows{a} p \con \knows{b} p \con \neg \distrib{}
    \common{} p, \knows{a} p, \knows{b} p, \neg \distrib{} \common{}
    p, \distrib{} p, p}$;

  $\G_1 = \set{\neg \common{} p, \knows{a} p, \knows{b} p, \neg
    \distrib{} \common{} p, \distrib{} p}$;

  $\D_2 = \set{\neg \common{} p, \knows{a} p, \knows{b} p, \neg
    \distrib{} \common{} p, \distrib{} p, p, \neg \knows{a} (p \con
    \common{} p)}$;

  $\D_3 = \set{\neg \common{} p, \knows{a} p, \knows{b} p, \neg
    \distrib{} \common{} p, \distrib{} p, p, \neg \knows{b} (p \con
    \common{} p)}$;

  $\G_2 = \set{\neg (p \con \common{} p), \knows{a} p, \neg \knows{a}
    (p \con \common{} p)}$;

  $\G_3 = \set{\neg \common{} p, \knows{a} p, \knows{b} p, \neg
    \distrib{} \common{} p, \distrib{} p, \neg \knows{a} (p \con
    \common{} p)}$;

  $\D_4 = \set{\neg p, \knows{a} p, \neg \knows{a} (p \con \common{}
    p), \distrib{} p, p}$;

  $\D_5 = \set{\neg \common{} p, \knows{a} p, \neg \knows{a} (p \con
    \common{} p), \distrib{} p, p}$;

  $\D_6 = \set{\neg \common{} p, \knows{a} p, \neg \knows{a} (p \con
    \common{} p), \distrib{} p, p, \neg \knows{b} (p \con \common{}
    p)}$;

  $\D_7 = \set{\neg \common{} p, \knows{a} p, \knows{b} p, \neg
    \distrib{} \common{} p, \distrib{} p, \neg \knows{a} (p \con
    \common{} p), p}$;

  $\D_8 = \set{\neg \common{} p, \knows{a} p, \knows{b} p, \neg
    \distrib{} \common{} p, \distrib{} p, \neg \knows{a} (p \con
    \common{} p), p, \\ \neg \knows{b} (p \con \common{} p)}$

  $\G_4 = \set{\neg \common{} p, \knows{a} p, \knows{b} p, \neg
    \distrib{} \common{} p, \distrib{} p, \neg \knows{b} (p \con
    \common{} p)}$

  $\G_5 = \set{\neg (p \con \common{} p), \knows{b} p, \neg \knows{b}
    (p \con \common{} p)}$

  $\D_9 = \set{\neg \common{} p, \knows{a} p, \knows{b} p, \neg
    \distrib{} \common{} p, \distrib{} p, \neg \knows{b} (p \con
    \common{} p), \neg \knows{a} (p \con \common{} p)}$

  $\D_{10} = \set{\neg \common{} p, \knows{a} p, \knows{b} p, \neg
    \distrib{} \common{} p, \distrib{} p, \neg \knows{b} (p \con
    \common{} p), p}$

  $\D_{11} = \set{\neg p, \knows{b} p, \neg \knows{b} (p \con
    \common{} p), \distrib{} p, p}$

  $\D_{12} = \set{\neg \common{} p, \knows{b} p, \neg \knows{b} (p
    \con \common{} p), \neg \knows{a} (p \con \common{} p), \distrib{}
    p, p}$

  $\D_{13} = \set{\neg \common{} p, \knows{b} p, \neg \knows{b} (p
    \con \common{} p), \distrib{} p, p}$

  $\G_6 = \set{\neg \common{} p, \knows{a} p, \knows{b} p, \neg
    \distrib{} \common{} p, \distrib{} p, \neg \knows{a} (p \con
    \common{} p), \neg \knows{b} (p \con \common{} p)}$

  $\D_{14} = \set{\neg \common{} p, \knows{a} p, \knows{b} p, \neg
    \distrib{} \common{} p, \distrib{} p, \neg \knows{a} (p \con
    \common{} p), \neg \knows{b} (p \con \common{} p), p}$

  $\G_7 = \set{\neg (p \con \common{} p)}$

  $\D_{15} = \set{\neg p}$;

  $\D_{16} = \set{\neg \common{} p, \neg \knows{a} (p \con
    \common{}p}$;

  $\D_{17} = \set{\neg \common{} p, \neg \knows{b} (p \con
    \common{}p}$
}

For lack of space, we do not depict the initial and final tableaux for
the input formula, but briefly describe what happens at the state
elimination stage.  States $\D_4$ and $\D_{11}$ get removed due to
\Rule{E1}, as they contain patent inconsistencies.  $\D_{14}$ gets
removed due to \Rule{E3}, since it contains an eventuality $\neg
\common{} p$ which is not realized in the tableau, as the rank of
$\D_{14}$ stabilizes at $\omega$, because it does not contain $\neg
p$, and is its only successor.  Then $\D_8$ and $\D_9$ get removed, as
their only successor along $\chi_0$, namely $\D_{14}$ has been
removed.  All other states remain in place; in particular, all of them
receive a finite rank, because from each of them one can reach the
state $\D_{15}$, which contains $\neg p$. The resultant graph encodes
all possible Hintikka structures for the input formula.
\end{example}

We note that our tableaux never close on account of all states
obtained from the initial prestate containing unfulfilled
eventualities (we omit the formal proof of this claim due to lack of
space). The rule \Rule{E3}, however, as can be seen from the example
above, eliminates from the tableau ``bad'' states, thus making our
tableau not only test a formula for satisfiability, but actually, for
every satisfiable formula $\theta$, produce a graph ``containing'' all
possible Hintikka structures for $\theta$ (i.e, whenever a node of the
graph is connected to several other nodes by arrows marked by the same
formula, these ``target'' nodes are not meant to be part of the same
MAEHS for $\theta$, but rather represent alternative ways of building a
MAEHS for $\theta$).

\section{Soundness and completeness}
\label{sec:soundness_and_completeness}

\subsection{Soundness}

The soundness of a tableau procedure amounts to claiming that if the
input formula $\theta$ is satisfiable, then the tableau for $\theta$
is open.  To establish soundness of the overall procedure, we prove a
series of lemmas that show that every rule is sound; the soundness of
the overall procedure will then easily follow. The proofs of the
following three lemmas are straightforward.

\begin{lemma}
  \label{lm:expansion}
  Let $\G$ be a prestate of $\tableau{P}^{\theta}$ such that
  \sat{M}{s}{\G} for some MAEM \mmodel{M} and $s \in
  \mmodel{M}$.  Then, \sat{M}{s}{\D} holds for at least one $\D \in
  \st{\G}$.
\end{lemma}

\begin{lemma}
  \label{lm:KR_sound}
  Let $\D \in S_0^{\theta}$ be such that \sat{M}{s}{\D} for some
  MAEM \mmodel{M} and $s \in \mmodel{M}$, and let $\neg
  \knows{a} \vp \in \D$.  Then, there exists $t \in \mmodel{M}$ such
  that $(s, t) \in \rel{R}_a$ and \sat{M}{t}{\set{\neg \vp} \union
    \crh{\knows{a} \psi}{\knows{a} \psi \in \D} \union \crh{\neg
      \knows{a} \psi}{\neg \knows{a} \psi \in \D}}.
\end{lemma}

\begin{lemma}
  \label{lm:DR_sound}
  Let $\D \in S_0^{\theta}$ be such that \sat{M}{s}{\D} for some
  MAEM \mmodel{M} and $s \in \mmodel{M}$, and let $\neg
  \distrib{} \vp \in \D$.  Then, there exists $t \in \mmodel{M}$ such
  that $(s, t) \in \rel{R}_D$ and \sat{M}{t}{\set{\neg \vp} \union
    \crh{\distrib{} \psi}{\distrib{} \psi \in \D} \union \crh{\neg
      \distrib{} \psi}{\neg \distrib{} \psi \in \D} \union
    \crh{\knows{a} \chi}{\knows{a} \chi \in \D, a \in \agents} \union
    \crh{\neg \knows{a} \chi}{\neg \knows{a} \chi \in \D, a \in \agents}}.
\end{lemma}

\begin{lemma}
  \label{lm:E3_sound}
  Let $\D \in S_0^{\theta}$ be such that \sat{M}{s}{\D} for some
  MAEM \mmodel{M} and $s \in \mmodel{M}$, and let $\neg
  \common{} \vp \in \D$.  Then, $\neg \common{} \vp$ is realized at
  $\D$ in $\tableau{T}_n^{\theta}$.
\end{lemma}

\begin{proof}
  As $\D$ is fully expanded, $\neg \knows{a} (\vp \con \common{} \vp)
  \in \D$ for some $a \in \agents$, and thus \sat{M}{s}{\neg \knows{a}
    (\vp \con \common{} \vp)}.  Therefore, there exists $s_1 \in
  \mmodel{M}$ such that $(s, s_1) \in \rel{R}_a$ and \sat{M}{s_1}{\neg
    (\vp \con \common{} \vp)}.  By construction of the tableau,
  \sat{M}{s_1}{\G} holds for the prestate $\G$ associated with $\neg
  \knows{a} (\vp \con \common{} \vp)$, i.e. such $\G$ that $\neg (\vp
  \con \common{} \vp) \in \G$.  Now, there exists $\D_1 \in \st{\G}$
  such that \sat{M}{s_1}{\D_1}.  Indeed, elements of $\st{\G}$ are
  full expansions of $\G$; clearly, $\G$ can be fully expanded in such
  a way that whenever we have to make a choice which of several
  formulae to include into $\D_1$ (say, for which $b \in \agents$ to
  add the formula $\neg \knows{b} (\vp \con \common{} \vp)$ if $\neg
  \common{} \vp \in \G$), we choose the one that is actually satisfied
  at $s_1$.  Now, as $\neg (\vp \con \common{} \vp) \in \G$, either
  \sat{M}{s_1}{\neg \vp} or \sat{M}{s_1}{\neg \common{} \vp}.  In the
  former case, we are done straight off, as then $\neg \vp \in \D_1$.
  In the latter case, as \sat{M}{s_1}{\neg \common{} \vp}, there
  exists a sequence of states $s_1, s_2, \ldots, s_m$ in \mmodel{M}
  such that for every $1 \leq i < m$, we have $(s_i, s_{i+1}) \in
  \rel{R}_b$ for some $b \in \agents$ and \sat{M}{s_m}{\neg \vp}.  By
  taking this sequence of states of \mmodel{M}, we can build, in the
  ``forcing choices'' style described above, a sequence of states
  $\D_1, \D_2, \ldots, \D_m \in S^{\theta}_n$ such that, for every $1
  \leq i < m$, we have $\D_i \stackrel{\neg \knows{b} (\vp \con
    \common{} \vp)}{\longrightarrow} \D_{i+1}$ for some $b \in
  \agents$, and $\neg \vp \in \D_m$.  The existence of the path $\D,
  \D_1, \ldots, \D_m$ implies that $\neg \common{} \vp$ is realized at
  $\D$ in $\tableau{T}^{\theta}_n$.
\end{proof}
\begin{theorem}[Soundness]
  \label{thr:soundness}
  If $\theta \in \lang$ is satisfiable in a MAEM, then
  $\tableau{T}^{\theta}$ is open.
\end{theorem}
\begin{proofsketch}
  Using the preceding lemmas, show by induction on the number of
  stages in the state elimination process that no satisfiable state
  can be eliminated due to \Rule{E1}--\Rule{E3}.  The claim then
  follows from lemma~\ref{lm:expansion}.
\end{proofsketch}

\subsection{Completeness}
\label{sec:completeness}

The completeness of a tableau procedure means that if the tableau for
a formula $\theta$ is open, then $\theta$ is satisfiable in a MAEM. By
making use of theorem~\ref{thr:sat_equal_hintikka}, it suffices to
show that an open tableau for $\theta$ can be turned into a MAEHS for
$\theta$. The construction of such a MAEHS is described in the
following lemma.

\begin{lemma}
  \label{lm:open_tableau_hintikka}
  If $\tableau{T}^{\theta}$ is open, then there exists a MAEHS for
  $\theta$.
\end{lemma}

\begin{proofsketch}
  Let $\tableau{T}^{\theta}$ be open.  The MAEHS \hintikka{H} for
  $\theta$ is built out of the so-called \emph{final tree components}.
  Each final tree component is a tree-like MAES with nodes labeled
  with states from $S^{\theta}$. Each component is associated with a
  state $\D \in S^{\theta}$ and an eventuality $\xi \in \ecl{\theta}$;
  such a component is denoted by $T_{\D,\xi}$.

  Now we describe how to build the final tree components. Let $\xi =
  \neg \common{} \vp \in \ecl{\theta}$ and $\D \in S^{\theta}$.  If
  $\xi \notin \D$, then $T_{\D,\xi}$ is a ``simple tree'' (i.e, one
  whose only inner node is the root) whose root is labeled with $\D$
  and that has exactly one leaf associated with each formula of the
  form $\neg \knows{a} \vp$ or $\neg \distrib{} \vp$ belonging to
  $\D$.  A leaf associated with formula $\chi$ is labeled by a state
  $\D' \in S^{\theta}$ such that in $\tableau{T}^{\theta}$ we have $\D
  \stackrel{\chi}{\longrightarrow} \D'$ (such a $\D'$
  exists---otherwise $\D$ would have been eliminated from the tableau
  due to \Rule{E2}).  To obtain a tree-like MAES, put $(s, t) \in
  \rel{R}_a$ if $s$ is labeled with $\D$, $t$ is labeled with $\D'$,
  and $\D \stackrel{\neg \knows{a} \vp}{\longrightarrow} \D'$ for some
  $\vp$; analogously, put $(s, t) \in \rel{R}_D$ if $s$ is labeled
  with $\D$, $t$ is labeled with $\D'$, and $\D \stackrel{\neg
    \distrib{} \vp}{\longrightarrow} \D'$ for some $\vp$.

  If, on the other hand, $\xi = \neg \common{} \vp \in \D$, then
  $T_{\D,\xi}$ is constructed as follows. Since $\neg \common{} \vp$
  is realized at $\D$ in $\tableau{T}^{\theta}$, there exists a
  sequence of states $\D = \D_0, \D_1, \ldots, \D_m$ in $S^{\theta}$
  such that $\neg \vp \in \D_m$ and for every $0 \leq i < m$, $\D
  \stackrel{\chi}{\longrightarrow} \D'$ holds for some $\chi$ of the
  form $\neg \knows{a} \vp$ or $\neg \distrib{} \vp$ (otherwise, it
  would have been eliminated due to \Rule{E3}).  Take this sequence
  and give to each $\D_i$ ($0 \leq i \leq m$) ``enough'' successors,
  as in the previous paragraph, and define the relations for this tree
  as prescribed therein.

  We are next going to stitch the above-defined $T_{\D,\xi}$'s
  together.  First, however, we note that if an eventuality $\xi'$
  belongs to $\D$ and is not realized inside some final tree component
  $T_{\D,\xi}$ (the realization in a final tree component is defined
  as in tableaux, with substituting $T_{\D,\xi}$ for
  $\tableau{T}^{\theta}_n$), then $\xi'$ belongs to every leaf of
  $T_{\D,\xi}$, and thus its realization is deferred---this is crucial
  to our ability to stitch $T_{\D, \xi}$'s up into a Hintikka
  structure.

  We now proceed as follows. First, we arrange all states of
  $\tableau{T}^{\theta}$ in a list $\D_0, \ldots, \D_{n-1}$ and all
  eventualities occurring in the states of $\tableau{T}^{\theta}$ in a
  list $\xi_{0}, \ldots, \xi_{m-1}$.  We then think of all final
  tree components as arranged in an $m$-by-$n$ grid whose rows are
  marked with the correspondingly numbered eventualities of
  $\tableau{T}^{\theta}$ and whose columns are marked with the
  correspondingly numbered states of $\tableau{T}^{\theta}$.  The
  final tree component at the intersection of the $i$th row and the
  $j$th column will be denoted by $T_{(i, j)}$. The building blocks
  for our MAEHS will all come from the grid. This MAEHS is built
  incrementally, so that at each stage of the construction we produce
  a structure realizing more and more eventualities.

  We start off with a final tree component that is uniquely determined
  by the input formula $\theta$, in the following way.  If $\theta$ is
  an eventuality, i.e., $\theta = \xi_p$ for some $0 \leq p < m$, then
  we start off with the component $T_{(p, q)}$ where, for
  definiteness, $q$ is the least number $< n$ such that $\theta \in
  \D_q$; as $\tableau{T}^{\theta}$ is open, such a $q$ exists.  If, on
  the other hand, $\theta$ is not an eventuality, then we start off
  with $T_{(0, q)}$, where $q$ is as described above.  Let's denote
  this initial structure by $\hintikka{H}_0$.

  Henceforth, we proceed as follows.  Informally, we think of the
  above list of eventualities as a queue of customers waiting to be
  served.  Unlike the usual queues, we do not necessarily start
  serving the queue from the first customer (if $\theta$ is an
  eventuality, then it gets served first; otherwise we start from the
  beginning of the queue), but then we follow the queue order, curving
  back to the beginning of the queue after having served its last
  eventuality, if we started in the middle.  Serving an eventuality
  $\xi$ amounts to appending to the leaves of the structure built thus
  far final tree components realizing $\xi$.  Thus, we keep track of
  what eventualities have already been served, take note of the one
  that was served the last, say $\xi_j$, and replace every leaf of the
  structure $\hintikka{H}_i$ constructed thus far with the final tree
  component $T_{i+1, ((j+1) \mod m)}$. The process continues until all
  eventualities have been served, at which point we have gone the full
  cycle through the queue.

  After that, the cycle is repeated, for as long as the queue remains
  non-empty. Alternatively, if we want to guarantee that the MAEHS we
  are building is going to be finite, the cycle is repeated with the
  following modification: whenever the component we are about to
  attach, say $T_{(i,j)}$, is already contained in our structure in
  the making, instead of replacing the leaf $t$ with that component,
  we connect every ``predecessor'' $s$ of $t$ to the root of
  $T_{(i,j)}$ with the relation connecting $s$ to $t$.  This modified
  version of the cycle is repeated until we come to a point when no
  more components get added---this is bound to happen in a finite
  number of steps as the number of $T_{\D, \xi}$'s is finite. It is
  now routine to check that the resultant structure $\hintikka{H}$ is
  a Hintikka structure, whose set of agents is the set of agents
  occurring in $\theta$.  By construction, it contains a node labeled
  with a set containing $\theta$.
\end{proofsketch}
\begin{theorem}[Completeness]
  \label{thr:completeness}
  Let $\theta \in \lang$ and let $\tableau{T}^{\theta}$ be open.
  Then, $\theta$ is satisfiable in a MAEM.
\end{theorem}
\begin{proof}
  Immediate from lemma~\ref{lm:open_tableau_hintikka} and
  theorem~\ref{thr:sat_equal_hintikka}.
\end{proof}

\section{Complexity of the procedure}

Let's denote the length of the input formula $\theta$ by $n$ and the
number of agents in the language by $k$. We assume that $k>1$,
otherwise we just deal with the modal logic \system{S5}.  The size of
the extended closure for $\theta$ (recall
definition~\ref{def:extended_closure}) is bounded from above by
\bigo{k^n}, as each $\common{}$ operator occurring in $\theta$
requires $k$ formulas to be added to the extended closure.

The examination of the procedure shows that the longest path to any
state of the pretableau we create at the construction phase from the
initial prestate (i.e., the one containing the input formula $\theta$)
is bound by the number of nested ``diamond'' modalities (such as $\neg
\knows{a}$) in $\theta$ plus 1.  From any given state or prestate we
can create at most \bigo{k^n} \linebreak (pre-)states, hence the whole
number of nodes we create is in \bigo{k^{n^2}}.  Thus, the
construction phase can be done in time \bigo{k^{n^2}}.

At the prestate elimination phase, we delete at most $\bigo{k^{n^2}}$
states and for each prestate redirect at most $\bigo{k^n}$ arrows,
which takes within \bigo{k^{n^2}} steps.

At the state elimination stage, we first apply \Rule{E1} to
\bigo{k^{n^2}} states, which can be done in \bigo{k^{(2n + n^2)}}
steps.  After that, we embark on the dovetailed application of
\Rule{E2} and \Rule{E3}.  We proceed in circles, whose number is bound
by \bigo{k^{n^2}}, as at each iteration we remove at least one state.
During each cycle, we carry out \bigo{k^n} times (the upper bound on
the number of eventualities) the following procedure: fist, we apply
\Rule{E2} to all states, which can be done in time $\bigo{k^{(n +
    n^2)}}$, and then apply \Rule{E3} to the pending eventuality.  The
latter procedure is carried out by computing a rank of each state of
the tableau with respect to the pending eventuality.  The number of
rank updates is bound by \bigo{k^{n^2}}, each update requiring
$\bigo{k^{(n + n^2)}}$ steps, as for each state $\D$ we check the
ranks of the targets of outgoing arrows marked by formulae in $\D$.
Thus, the whole state elimination phase can be carried out in
$\bigo{k^{2n^2}}$ steps.

We conclude that the whole procedure can be carried out in
$\bigo{k^{2n^2}}$ steps, where $n$ is the size of the input formula.
It follows that \maelcd-satisfiability is in \cclass{ExpTime}, which
together with the result from~\cite{HM92} implies that
\maelcd-satisfiability is \cclass{ExpTime}-complete.

\section{Concluding remarks}
\label{sec:concluding}

We have developed a sound, complete, and complexity-optimal
incremental-tableau-based decision procedure for the multi-agent
epistemic logic \maelcd. We claim that this style of tableau is of
immediate practical use, both by human and computerized execution.  It
is more efficient (within the theoretically established complexity
bounds) and more modular and adaptable than the top-down tableaux of
the type developed (for a fragment of the logic not including the
$\distrib{}$ operator) in~\cite{HM92}. In particular, the tableaux
presented lends itself to an extension to the full multi-agent
epistemic logic, with modal operators of common and distributed
knowledge for all coalitions of agents, and well as to a combination
with the similar style tableaux developed for the Alternating-time
temporal logic $\ATL$ developed in \cite{GorSh08}, which are going to
be the subject of our subsequent work.

\vspace{0.2cm}

\noindent \textbf{Acknowledgments} This research was supported by a
research grant of the National Research Foundation of South Africa and
was done during the second author's post-doctoral fellowship at the
University of the Witwatersrand, funded by the Claude Harris Leon
Foundation---we gratefully acknowledge the financial support from these
institutions. We also acknowledge the anonymous referees whose remarks
helped to improve our presentation.

\end{document}